\documentclass[aps,pra,footinbib, showpacs,showkeys,longbibliography]{revtex4-2}
\usepackage [latin1]{inputenc}
\usepackage{graphicx}
\usepackage{indentfirst}
\usepackage{physics}
\usepackage{braket}
\usepackage{float}
\usepackage{amsmath}
\usepackage{amssymb}
\usepackage{verbatim}
\usepackage{wasysym}
\usepackage{epstopdf}
\usepackage{CJK}
\usepackage{esint}
\usepackage{color}
\usepackage{xcolor}
\usepackage{epsfig}
\usepackage{subfigure}
\usepackage{amsfonts}
\usepackage{footmisc}
\usepackage{scrextend}
\usepackage{multirow}
\usepackage{mathtools}
\newcommand\numberthis{\addtocounter{equation}{1}\tag{\theequation}}

\usepackage{xr-hyper}
\usepackage[hyperfootnotes=false]{hyperref}

\usepackage{amsmath}

\usepackage[english]{babel}
\usepackage{url}
\usepackage{bm}
\definecolor{darkblue}{rgb}{0,0,0.5}
\hypersetup{
colorlinks=true,
linkcolor=black,
filecolor=black,
citecolor=darkblue,
urlcolor=darkblue,
}

\begin{document}

\title{Continuous-variable measurement device independent\\ quantum conferencing with post-selection}
\author{Alasdair I.~Fletcher}
\author{Stefano Pirandola}
\affiliation{Department of Computer Science, University of York, York YO10 5GH, United Kingdom}

\begin{abstract}
A continuous variable (CV), measurement device independent (MDI) quantum key distribution (QKD) protocol is analyzed, enabling three parties to connect for quantum conferencing. We utilise a generalised Bell detection at an untrusted relay and a postselection procedure, in which distant parties reconcile on the signs of the displacements of the quadratures of their prepared coherent states. We derive the rate of the protocol under a collective pure-loss attack, demonstrating improved rate-distance performance compared to the equivalent non-post-selected protocol. In the symmetric configuration in which all the parties lie the same distance from the relay, we find a positive key rate over 6 km. Such postselection techniques can be used to improve the rate of multi-party quantum conferencing protocols at longer distances  at the cost of reduced performance at shorter distances.
\end{abstract}
\maketitle
\section{Introduction}
Quantum Key Distribution (QKD) promises provably secure communication \cite{Pirandola2020} based on fundamental physical principles. Relying on the inability to clone arbitrary quantum states \cite{Wootters1982} and by utilising non-orthogonal states or entanglement \cite{Ekert1991}, two distant parties are able to agree symmetric cryptographic keys, secure against any attack possible within the laws of quantum mechanics. The technology has rapidly matured, advancing from the first proposed protocols based on transmission of discrete single qubit states \cite{C.H.BennettandG.Brassard1984,Bennett1992} and proof of principle of experiments to practical deployments over long distances \cite{Stucki2009,Lucamarini2018,Pittaluga21} and networks and network protocols enabling multiple users to communicate securely across metropolitan sized areas and beyond \cite{Joshi2020,Dynes2019,Solomons2021}.

However, whilst QKD offers ultimate security against channel attacks, its practical implementation remains challenging. Many approaches require trusted experimental devices and detectors and therefore suffer from the possibility of so-called \textit{side-channel attacks} against such devices. Fully Device-Independent approaches to QKD are possible, which entirely eliminate such attacks \cite{Barrett2005,Schwonnek2021,Pironio2009} but these are practically limited by low rates and poor distance scaling. Instead Measurement Device Independent (MDI) QKD \cite{Braunstein2012,Lo2012} provides a middle ground, relaxing the assumptions on the protocol by having distant parties send states to a central relay detector which may be controlled by an Eavesdropper (Eve). Malicious behaviour by Eve may be detected by the parties in the reconciliation and parameter estimation stage of the protocol.

Moreover, point-to-point quantum communications are known to be inherently distance limited by the PLOB bound \cite{Pirandola2017} expressed by the formula $\mathcal{C}=-\log_2(1-\eta)$ with the transmissivity $\eta$ decaying exponentially with distance. Continuous variable (CV) QKD protocols are able to reach rates approaching the PLOB bound, outperforming discrete state protocols; furthermore their experimental implementation is more straightforward \cite{Pirandola2020,Laudenbach2018}. Naively, there was thought to be a $3$db (corresponding to $\eta=\frac{1}{2}$) loss-limit on CV QKD, however this has since been exceeded with reverse reconciliation and postselection techniques.  Postselection techniques rely on the fact that even beyond $3$db loss there are regions in parameter space in which the rate remains positive~\cite{Silberhorn2002}. By announcing the absolute values of the quadratures of their prepared coherent states the two end parties are able to select only these regions, reconciling the signs of their quadratures into a key with a positive rate even beyond $3$db. This approach was also implemented experimentally \cite{Symul2007}.

Such post selection techniques have recently exploited to extend the distance of two-party CV MDI QKD \cite{Wilkinson2020} and in this work we demonstrate that the same approach can be utilised to increase the effective range at which CV MDI quantum conferencing can occur by utilising a generalised Bell detection introduced in \cite{Ottaviani2019}. Whilst we are restricted by the need to perform numerical integration in large number of dimensions to consider only three parties and pure loss attacks, the protocol presented here is in principle readily extended to $N$ users and entangling cloner attacks. 

The structure of the paper is as follows: In Sec. \ref{sec:protocol} we introduce the protocol and explain the structure of the detector; Sec. \ref{sec:rate} explains how the rate of the protocol is calculated; Sec. \ref{sec:results} provides results and Sec. \ref{sec:conclusion} is for conclusions.

\section{Protocol and Detector}
\label{sec:protocol}
In this paper we consider the case of three users undertaking quantum conferencing. The three parties: Alice, Bob and Charlie individually prepare Gaussian modulated coherent states. Each party individually has access to an independent zero-mean Gaussian distribution with standard deviations $\sigma_A,\sigma_B,\sigma_C$ respectively. Each party then draws two independent values from their respective distributions for the value of the $q$ and $p$ quadratures of their coherent state. They encode the absolute values in the variables $\mathbb{Q}_i$ and $\mathbb{P}_i$  respectively and the signs in $\kappa_i$ and $\kappa_i'$. Thus they prepare coherent states of the form:

\begin{equation}
\ket{\alpha_i}=\ket{\frac{1}{2}(\kappa_i\mathbb{Q}_i+\kappa_i'\mathbb{P}_i)} \ \ \mathrm{for} \  i=A,B,C \ .
\end{equation}

Each state is sent through a lossy channel to the detector which may be attacked by an eavesdropper (Eve). This is modelled as a beamsplitter attack in which Eve inserts a beamsplitter into each channel, storing the outputs in a quantum memory. In a pure loss attack, Eve does not actively inject any state at the beamsplitter and thus each coherent state is instead mixed with the vacuum state $\ket{0}$.

\begin{figure}[h]
    \centering
    \includegraphics[width=0.7\linewidth]{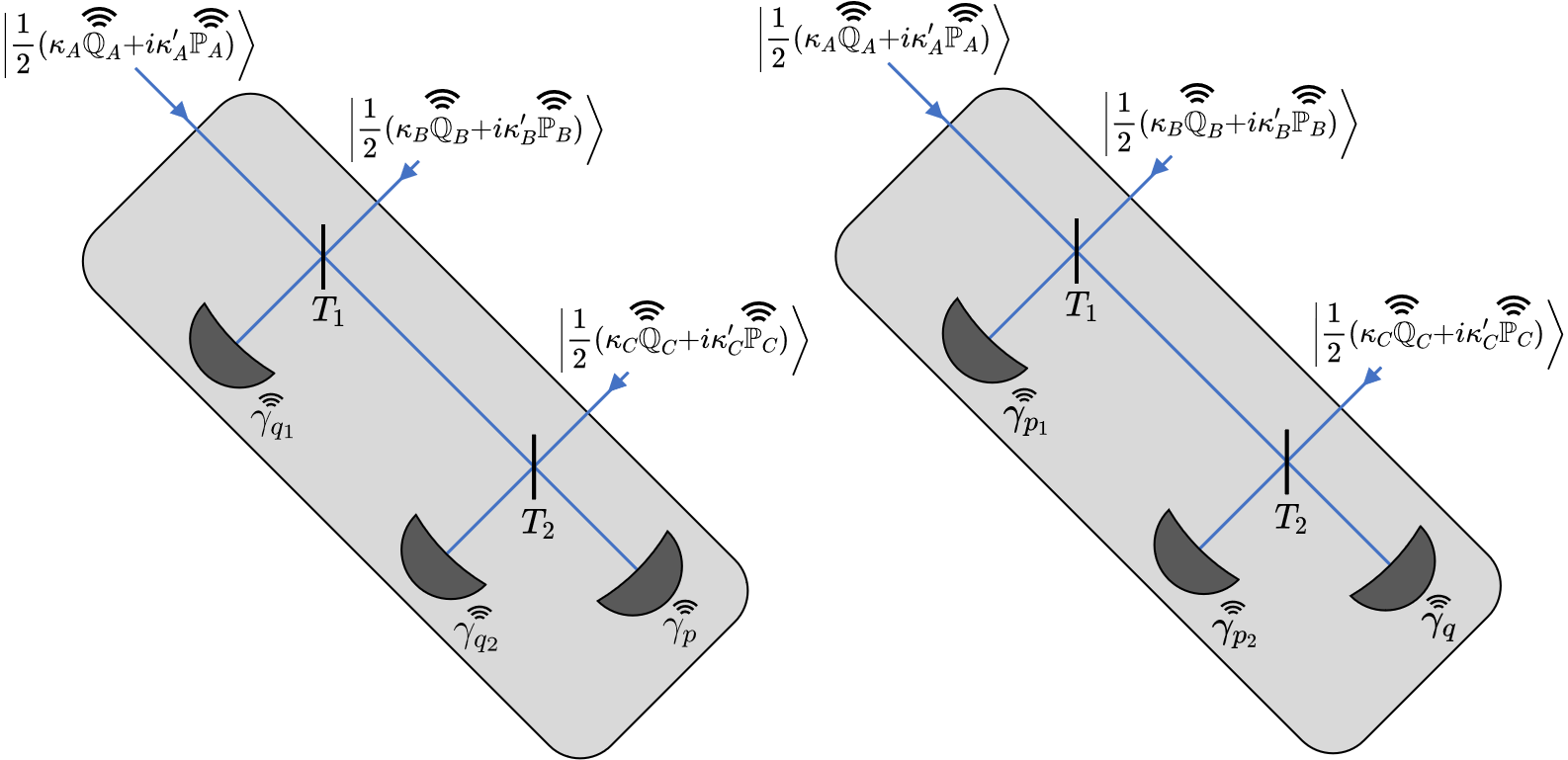}
    \caption{Structure of the detector, demonstrating the two possible orientations. Input modes are mixed by two beamsplitters with transmissivities $T_1=\frac{1}{2}$ and $T_2=\frac{2}{3}$. In the first configuration (pictured left) the states undergo two $q$ homodyne detections and one $p$ homodyne detection. The parties will attempt reconciliation between $\kappa_A',\kappa_B',\kappa_C'$. In the second orientation (pictured right) the states undergo two $p$ homodyne detections and one $q$ homodyne detection. In this case the parties attempt reconciliation on $\kappa_A,\kappa_B,\kappa_C$. }
    \label{fig: detector config}
\end{figure}

The structure of the detector is illustrated in Fig \ref{fig: detector config} and was devised in \cite{Ottaviani2019} to perform a generalised Bell detection on the incoming coherent states. It is comprised of a cascade of beamsplitters, each having transmissivity $T_i=\frac{i}{i+1}$. In the case of three parties, which we consider, this corresponds to $T_1=1/2$ and $T_2=2/3$. The beamsplitters are followed by $N-1 ~ q ~ (p)$ homodyne detections and a final homodyne detection in the $p ~ (q)$ quadrature and the results of all the measurements are publicly broadcast. The two possible configurations are switched between randomly and are announced at the end of the protocol. At this point each party reveals their values of $\mathbb{Q}_i$ and $\mathbb{P}_i$, and publicly broadcasts them to every other user. Operated correctly, in the entanglement based representation the detector has the effect of projecting the Alice-Bob-Charlie state into a symmetric state with GHZ-like correlations between each parties state \cite{Ottaviani2019}. 

\begin{figure}
    \centering
    \includegraphics[width=0.7\linewidth]{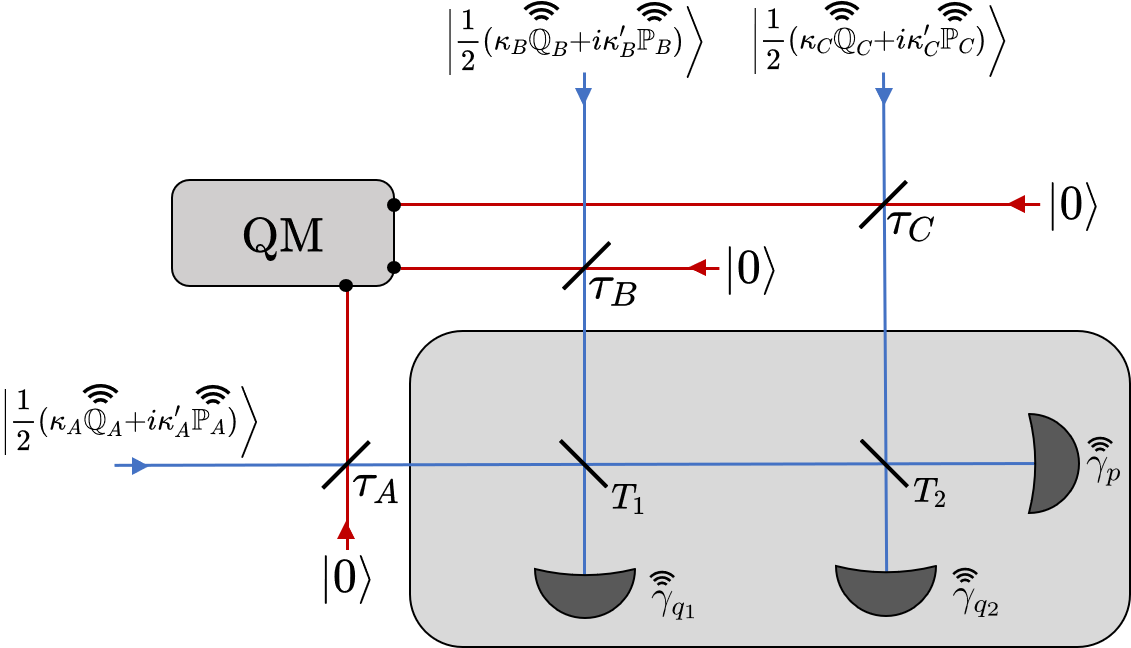}
    \caption{Operation of the detector under a collective pure loss attack. Eve attacks each of the incoming channels by inserting beamsplitters with transmissivities $\tau_A, \tau_B, \tau_C$, which combine the incoming signals with vacuum states $\ket{0}$. Eve stores her output modes in a quantum memory (QM). The remaining modes are mixed in the cascade of beamsplitters and then undergo homodyne detection. The results of the homodyne detections $\gamma_{q_1},\gamma_{q_2},\gamma_{p}$ are publicly announced. Alice, Bob and Charlie also publicly announce the absolute values of the quadratures of their prepared coherent states $\mathbb{Q}_A,\mathbb{Q}_B,\mathbb{Q}_C,\mathbb{P}_A,\mathbb{P}_B,\mathbb{P}_C$. In this configuration the parties attempt to reconcile their values of $\kappa_A',\kappa_B',\kappa_C'$.}
    \label{pure-loss-attack}
\end{figure}

\section{Rate}
\label{sec:rate}
We first sketch the method used to determine the rate. At the end of the protocol the parties perform pairwise reconciliation between $\kappa_A,\kappa_B,\kappa_C$ or $\kappa_A',\kappa_B',\kappa_C'$ depending on the orientation of the detector. In the asymptotic limit of a large number of uses the rate of the protocol is given by:
\begin{equation}
R_{ij}=I_{ij}-\chi
\end{equation}
where  $I_{ij}$ is the binary mutual information between the sign variables $\kappa_i$ and $\kappa_j$ or $\kappa'_i$ and $\kappa_j'$. $\chi$ is the Holevo information. The mutual information can therefore be found by utilising Bayes' Theorem and the distribution of measurement outcomes as detailed in Sec. \ref{sec:mutual}. The Holevo information is calculated by carefully considering Eve's state at the end of the protocol has explained in Sec. \ref{sec:holevo}. Additionally, since we ultimately wish to perform postselection to increase the performance of the protocol we work with \textit{single-point} versions of the above quantities $\tilde{I}_{ij}$  and $\chi$ which are the values conditioned upon the quadratures and measurement outcome. To this end we start by considering the initial covariance matrix of the Alice-Bob-Charlie-Eve system, which is given by:
\begin{equation}
\mathbf{V}_{ABCE}=\mathbf{I}_A\oplus\mathbf{I}_B\oplus\mathbf{I}_C\oplus\mathbf{V}_{E}
\end{equation}
where $\mathbf{I}$ is the two-by-two identity matrix and for a pure loss attack $\mathbf{V}_{E}=\mathbf{I}\oplus\mathbf{I}\oplus\mathbf{I}$.
%
The mean value of the Alice-Bob-Charlie system is:
\begin{equation}
\mathbf{\bar{x}}_{ABC}=(\kappa_A\mathbb{Q}_A,\kappa_A'\mathbb{P}_A,\kappa_B\mathbb{Q}_B,\kappa_B'\mathbb{P}_B,\kappa_C\mathbb{Q}_C,\kappa_C'\mathbb{P}_C)^T \  .
\end{equation}
The mean value of Eve's system is zero. After propagation through the detector's array of beamsplitters and the homodyne detections, the distribution of measurement outcomes is given by:
\begin{equation}
p(\gamma_p|\kappa_A',\kappa_B',\kappa_C',\mathbb{P}_A,\mathbb{P}_B,\mathbb{P}_C)=\frac{1}{\sqrt{2\pi v_p}}\mathrm{exp}\bigg(\frac{-(\gamma_p-\bar{p})^2}{2} \bigg)
\label{eq: dist}
\end{equation}
where

\begin{align*}
\bar{p}=\sqrt{T_1T_2\tau_A}\ \kappa_A' \mathbb{P}_A+\sqrt{(1-T_1)T_2\tau_B} \ \kappa_B' \mathbb{P}_B&  + \sqrt{(1-T_2)\tau_C} \ \kappa_C'\mathbb{P}_C \ . \numberthis 
\end{align*}
We have implicitly removed the conditioning on the modulus and absolute value of the $q$ quadratures from the notation as there is no dependence upon them. 
Similarly for the opposite detector configuration:
\begin{equation}
p(\gamma_q|\kappa_A,\kappa_B,\kappa_C,\mathbb{Q}_A,\mathbb{Q}_B,\mathbb{Q}_C)=\frac{1}{\sqrt{2\pi v_q}}\mathrm{exp}\bigg( \frac{-(\gamma_{q}-\bar{q})^2}{2}\bigg)
\end{equation}
where
\begin{equation}
\bar{q}=\sqrt{T_1T_2\tau_A}\ \kappa_A \mathbb{Q}_A+\sqrt{(1-T_1)T_2\tau_B} \ \kappa_B \mathbb{Q}_B  + \sqrt{(1-T_2)\tau_C} \ \kappa_C\mathbb{Q}_C  \ . 
\end{equation}
Finally, we have implicitly assumed throughout that the homodyne detectors have perfect efficiency.

\subsection{Mutual Information}
\label{sec:mutual}
We first introduce the following compact notation $\boldsymbol{\kappa'}=(\kappa_A',\kappa_B',\kappa_C')$; $\pmb{\mathbb{P}}=(\mathbb{P}_A,\mathbb{P}_B,\mathbb{P}_C)$, $\boldsymbol{\kappa'}_{\char`\\ A}=(\kappa_B',\kappa_C')$ which simplifies the following expressions. Let us recall the definition of the single point mutual information between the two binary variables $\kappa_i'$ and $\kappa_j'$. This is clearly just the mutual information conditioned on the announced variables $\gamma_p$ and $\pmb{\mathbb{P}}$:
\begin{equation}
\label{eq:single point}
\tilde{I}_{ij}=H_{\kappa_{i}'|\pmb{\mathbb{P}},\gamma_p}-\sum_{\kappa_{j}'}p(\kappa_{j}'|~\pmb{\mathbb{P}},\gamma_p)~H_{\kappa_{i}'|\kappa_{j}'\pmb{\mathbb{P}},\gamma_p} \ ,
\end{equation}
where $H$ is the binary entropy so that:
\begin{equation}
H_{\kappa_{i}'|\pmb{\mathbb{P}},\gamma_p}=-p(\kappa_{i}'|\pmb{\mathbb{P}},\gamma_p)\log_2\big(p(\kappa_{i}'|\pmb{\mathbb{P}},\gamma_p)\big)-\big(1-p(\kappa_{i}'|\pmb{\mathbb{P}},\gamma_p)\big)\log_2\big(1-p(\kappa_{i}'|\pmb{\mathbb{P}},\gamma_p)\big) \ ,
\end{equation}
and
\begin{equation}
H_{\kappa_{i}'|\kappa_{j}',\pmb{\mathbb{P}},\gamma_p}=-p(\kappa_{i}'|\kappa_{j}',\pmb{\mathbb{P}},\gamma_p)\log_2(p(\kappa_{i}'|\kappa_{j}',\pmb{\mathbb{P}},\gamma_p))-\Big(1-p(\kappa_{i}'|\kappa_j',\pmb{\mathbb{P}},\gamma_p)\Big)\log_2\Big(1-p(\kappa_{i}'|\kappa_j',\pmb{\mathbb{P}},\gamma_p)\Big) \ .
\end{equation}
From the symmetry of the detector we have $I_{AB}=I_{AC}=I_{BC}$ and for simplicity we consider only $I_{AB}$ from this point onwards. Using Eq.~(\ref{eq: dist}) and Bayes' theorem  we first calculate the probability of positive and negative values for $\kappa_A'$ conditioned on $\kappa_B'$,$\kappa_C'$, the magnitudes of the $p$ quadratures $\pmb{\mathbb{P}}$ and the measurement outcome $\gamma_p$:



\begin{equation}
p(\kappa_A'|\boldsymbol{\kappa'}_{\char`\\ A},\pmb{\mathbb{P}},\gamma_p)=\frac{p(\gamma_p|\boldsymbol{\kappa}',\pmb{\mathbb{P}})~p(\kappa'_A|\boldsymbol{\kappa'}_{\char`\\ A},\pmb{\mathbb{P}})}{p(\gamma_p|\boldsymbol{\kappa'}_{\char`\\ A},\pmb{\mathbb{P}})}
\end{equation}
Noting that,
\begin{equation}
p(\gamma_p|\boldsymbol{\kappa}'_{\char`\\ A},\pmb{\mathbb{P}})=\sum_{\kappa'_A}p(\gamma_p|\boldsymbol{\kappa}',\pmb{\mathbb{P}})~p(\kappa'_A|\boldsymbol{\kappa}'_{\char`\\ A},\pmb{\mathbb{P}})
\end{equation}
and $p(\kappa'_A|\boldsymbol{\kappa'}_{\char`\\ A},\pmb{\mathbb{P}})=1/2$ we reach:
\begin{equation}
p(\kappa_A'|\boldsymbol{\kappa'}_{\char`\\ A},\pmb{\mathbb{P}},\gamma_p)=\frac{p(\gamma_p|\boldsymbol{\kappa}',\pmb{\mathbb{P}})}{\sum_{\kappa_A'}p(\gamma_p|\boldsymbol{\kappa}',\pmb{\mathbb{P}})} .
\end{equation}
We may then remove the conditioning on $\kappa'_C$ to find $p(\kappa_A'|\kappa_B', \pmb{\mathbb{P}},\gamma_p)$ for the second term in the single point mutual information.
\begin{equation}
p(\kappa_A'|\kappa_B' , \pmb{\mathbb{P}},\gamma_p)=\sum_{\kappa_C'}p(\kappa_A'|\boldsymbol{\kappa'}_{\char`\\ A},\pmb{\mathbb{P}}\gamma_p)~p(\boldsymbol{\kappa'}_{\char`\\ B}|\kappa_B',\pmb{\mathbb{P}}),
\end{equation}
so that we may write
\begin{equation}
p(\kappa_A'|\kappa_B' , \pmb{\mathbb{P}})=\frac{\sum_{\kappa_C'}p(\gamma_p|\boldsymbol{\kappa}',\pmb{\mathbb{P}})}{\sum_{\kappa_A'\kappa_C'}p(\gamma_p|\boldsymbol{\kappa},'\pmb{\mathbb{P}})}.
\end{equation}
Similarly to further remove the dependence from $\kappa'_B$:
\begin{equation}
p(\kappa'_A|\pmb{\mathbb{P}},\gamma_p)=\frac{\sum_{\kappa'_B\kappa_C'}p(\gamma_p|\boldsymbol{\kappa}',\pmb{\mathbb{P}})}{\sum_{\kappa'_A\kappa'_B\kappa_C'}p(\gamma_p|\boldsymbol{\kappa}',\pmb{\mathbb{P}})}.
\end{equation}
By the same approach we can also find $p(\kappa_B'|\pmb{\mathbb{P}},\gamma_p)$, enabling the sum in Eq.~(\ref{eq:single point}) to be taken. Finally in order to take the integral over the single point mutual information we require the probability of all the variables
\begin{equation}
p(\gamma_p,\pmb{\mathbb{P}})=\sum_{\boldsymbol{\kappa'}} p(\gamma_p|\boldsymbol{\kappa'}\pmb{\mathbb{P}})~p(\kappa_A' \mathbb{P}_A)~p(\kappa_B' \mathbb{P}_B)~p(\kappa_C' \mathbb{P}_C).
\end{equation}

\subsection{Holevo Bound}
\label{sec:holevo}
At the end of the protocol Eve is left with the state $\hat{\rho}_{\mathfrak{E}|\pmb{\mathbb{P}},\gamma_p}$ which is her total state conditioned on the announced absolute values of the $p$ quadratures $\pmb{\mathbb{P}}$ and the measurement outcome $\gamma_p$. This state is a convex combination of pure Gaussian states corresponding to given values of $\kappa_A',\kappa_B',\kappa_C'$ and hence Eve's total state may be written:
\begin{equation}
\hat{\rho}_{\mathfrak{E}|\pmb{\mathbb{P}},\gamma_p}=\sum_{\boldsymbol{\kappa'}}p(\boldsymbol{\kappa'}|\pmb{\mathbb{P}},\gamma_p)~\hat{\rho}_{\mathfrak{E}|\boldsymbol{\kappa'},\pmb{\mathbb{P}},\gamma_p}.
\end{equation}
It is important to note that whilst the conditional states, $\hat{\rho}_{\mathfrak{E}|\boldsymbol{\kappa'},\pmb{\mathbb{P}},\gamma_p}$, are pure and Gaussian the total state, $\hat{\rho}_{\mathfrak{E}|\pmb{\mathbb{P}},\gamma_p}$ is not, which complicates our analysis. Nonetheless, assuming that Eve performs a collective attack on the protocol the relevant quantity to calculate is the Holevo information $\chi$. We can again write this as a single point quantity in the following way.

\begin{equation}
\tilde{\chi}(\mathfrak{E}:\kappa_{i}'|~\pmb{\mathbb{P}},\gamma_p)=S(\hat{\rho}_{\mathfrak{E}|\pmb{\mathbb{P}},\gamma_p})-S(\hat{\rho}_{\mathfrak{E}|\kappa_{i}',\pmb{\mathbb{P}},\gamma_p})
\end{equation}
where $\tilde{\chi}(\mathfrak{E}:\kappa_{i}'|~\pmb{\mathbb{P}},\gamma_p)$ is the single point Holevo information and $S$ is the von Neumann entropy  which we recall is calculated from the eigenvalues $\{\lambda_i\}$ of a density matrix $\hat{\rho}$ by:
\begin{equation}
S(\hat{\rho})=-\sum_i \lambda_i\log_2(\lambda_i).
\end{equation}

First let us write the conditional states $\hat{\rho}_{\mathfrak{E}|\boldsymbol{\kappa'}\pmb{\mathbb{P}}\gamma_p}$ as:
\begin{equation}
\hat{\rho}_{\mathfrak{E}|\boldsymbol{\kappa'}\pmb{\mathbb{P}}\gamma_p}=\ket{\mathfrak{E}^{\pmb{\mathbb{P}},\gamma_p}_{\kappa_{A}'\kappa_{B}'\kappa_{C}'}}\bra{\mathfrak{E}^{\pmb{\mathbb{P}},\gamma_p}_{\kappa_{A}'\kappa_{B}'\kappa_{C}'}}
\end{equation}
We consider the matrix of overlaps $O$ of this state for all the combinations of $\kappa_A',\kappa_B',\kappa_C'$.
\begin{align}
\label{eq:O}
O=
&\begin{pmatrix}
1 & C & B & BC & A & AC & AB & ABC \\
C & 1 & BC & B & AC & A & ABC & AB \\
B & BC & 1 & C & AB & ABC & A & AC \\
BC & B & C & 1 & ABC & AB & AC & A \\
A & AC & AB & ABC & 1 & C & B & BC \\
AC & A & ABC & AB & C & 1 & BC & B \\
AB & ABC & A & AC & B & BC & 1 & C \\
ABC & AB & AC & A & BC & B & C & 1 
\end{pmatrix} 
\begin{matrix}
(-1 & -1 &-1) \\ (-1 &-1 & 1) \\ (-1&1&-1) \\ (-1&1&1) \\ (1&-1&-1) \\ (1&-1&1) \\ (1&1-&1) \\ (1&1&1)
\end{matrix}
\end{align}
The values in the far column denote the row values of $\kappa_A',\kappa_B',\kappa_C'$. The columns may be similarly labelled. $O$ is clearly separable as:
\begin{align}
O=
\begin{pmatrix}
1 & A \\
A & 1 
\end{pmatrix}
\otimes
\begin{pmatrix}
1 & B \\
B & 1
\end{pmatrix}
\otimes
\begin{pmatrix}
1 & C \\
C & 1
\end{pmatrix}
\end{align}
which implies:
\begin{equation}
\ket{\mathfrak{E}^{\pmb{\mathbb{P}},\gamma_p}_{\kappa_{A}'\kappa_{B}'\kappa_{C}'}}=\ket{\mathfrak{E'}_{\kappa_{A}'}^{\pmb{\mathbb{P}},\gamma_p}} \otimes \ket{\mathfrak{E}^{\pmb{\mathbb{P}},\gamma_p}_{\kappa_{B}'}}\otimes\ket{\mathfrak{E'}_{\kappa_{C}'}^{\pmb{\mathbb{P}},\gamma_p}}.
\end{equation}
Each of these states lies in a two-dimensional Hilbert space. Using $x$ to index the parties $A,B,C$ we may expand the states as:
\begin{align}
\ket{\mathfrak{E}_{\kappa_{i}=-1}^{\pmb{\mathbb{P}},\gamma_p}}=c_{0}\ket{\Phi^{({x})}_{0}}+c_{1}\ket{\Phi^{(x)}_{1}}
\\\ket{\mathfrak{E}_{\kappa_{i}=1}^{\pmb{\mathbb{P}},\gamma_p}}=c_{0}\ket{\Phi_{0}^{(x)}}-c_{1}\ket{\Phi_{1}^{(x)}}
\end{align}
and find the following relation for the coefficients:
\begin{align}
|c^{(x)}_{0}|^2=\frac{1}{2}(1+X)
\\|c^{(x)}_{1}|^2=\frac{1}{2}(1-X)
\end{align}
where $X$ labels the corresponding values $A,B,C$ from Eq.~(\ref{eq:O}). For two Gaussian states with the same covariance matrix $\mathbf{V}$ and mean values $\bar{\mathbf{x}}_{1}$ and $\bar{\mathbf{x}}_{2}$ the following relation holds \cite{Banchi2015}:
\begin{equation}
\mathrm{Tr}(\hat{\rho}_{1}\hat{\rho}_{2})=\mathrm{exp}\bigg(-\frac{1}{4}(\mathbf{\bar{x}}_1-\mathbf{\bar{x}}_2)\mathbf{V}^{-1}(\mathbf{\bar{x}}_1-\mathbf{\bar{x}}_2)\bigg)
\end{equation}
which we use to calculate
\begin{align}
A=\braket{\mathfrak{E}_{\kappa_{A}=-1}^{\pmb{\mathbb{P}},\gamma_p}|\mathfrak{E}_{\kappa_{A}=1}^{\pmb{\mathbb{P}},\gamma_p}},
\\
B=\braket{\mathfrak{E}_{\kappa_{B}=-1}^{\pmb{\mathbb{P}},\gamma_p}|\mathfrak{E}_{\kappa_{B}=1}^{\pmb{\mathbb{P}},\gamma_p}},
\\
C=\braket{\mathfrak{E}_{\kappa_{C}=-1}^{\pmb{\mathbb{P}},\gamma_p}|\mathfrak{E}_{\kappa_{C}=1}^{\pmb{\mathbb{P}},\gamma_p}}.
\end{align}
We are now able to give $\hat{\rho}_{\mathfrak{E}|\pmb{\mathbb{P}},\gamma_p}$ in the $\{\ket{\Phi_{0}^{(A)}},\ket{\Phi_{1}^{(A)}}\}\otimes \{\ket{\Phi_{0}^{(B)}},\ket{\Phi_{1}^{(B)}}\}\otimes \{\ket{\Phi_{0}^{(C)}},\ket{\Phi_{1}^{(C)}}\}$ basis. Describing the row position with the binary string $(i,j,k)$ and similarly the column position with $(i',j',k')$ each component of the density matrix can be calculated by:

\begin{equation}
(\hat{\rho}_{\mathfrak{E}|\pmb{\mathbb{P}}\gamma_p})_{(ijk)(i'j'k')}=\sum_{\boldsymbol{\kappa'}}p(\boldsymbol{\kappa'}|\pmb{\mathbb{P}},\gamma_p)\braket{\Phi_{i}^{(A)}|\mathfrak{E}^{\pmb{\mathbb{P}},\gamma_p}_{\kappa_A}}\braket{\mathfrak{E}^{\pmb{\mathbb{P}},\gamma_p}_{\kappa_A}|\Phi_{i'}^{(A)}}\braket{\Phi_{j}^{(B)}|\mathfrak{E}^{\pmb{\mathbb{P}},\gamma_p}_{\kappa_B}}\braket{\mathfrak{E}^{\pmb{\mathbb{P}},\gamma_p}_{\kappa_B}|\Phi_{j'}^{(2)}}\braket{\Phi_{k}^{(C)}|\mathfrak{E}^{\pmb{\mathbb{P}},\gamma_p}_{\kappa_C'}}\braket{\mathfrak{E}^{\pmb{\mathbb{P}},\gamma_p}_{\kappa_C'}|\Phi_{k'}^{(C)}}.
\end{equation}
By calculating the following inner products:

\begin{align}
\braket{\Phi^{(x)}_0|\mathfrak{E}^{\pmb{\mathbb{P}},\gamma_p}_{\kappa_x=-1}}=c^{(x)}_0 \\
\braket{\Phi^{(i)}_0|\mathfrak{E}^{\pmb{\mathbb{P}},\gamma_p}_{\kappa_x=1}}=c^{(x)}_0 \\
\braket{\Phi^{(i)}_1|\mathfrak{E}^{\pmb{\mathbb{P}},\gamma_p}_{\kappa_x=-1}}=c^{(x)}_1 \\
\braket{\Phi^{(i)}_1|\mathfrak{E}^{\pmb{\mathbb{P}},\gamma_p}_{\kappa_x=1}}=-c^{(x)}_1 
\end{align}
we can therefore immediately find the diagonal components of the density matrix:
\begin{align}
(\hat{\rho}_{\mathfrak{E}|\pmb{\mathbb{P}},\gamma_p})_{(ijk)(ijk)}=|c_{i}^{(A)}|^2 \ |c_{j}^{(B)}|^2 \ |c_{k}^{(C)}|^2.
\end{align}
The off diagonal terms are given by:
\begin{equation}
(\hat{\rho}_{\mathfrak{E}|\pmb{\mathbb{P}},\gamma_p})_{(ijk)(i'j'k')}=c_{i}^{(A)}\big(c_{i'}^{(A)}\big)^*c_{j}^{(B)}\big(c_{j'}^{(B)}\big)^*c_{k}^{(C)}\big(c_{k'}^{(C)}\big)^*\Lambda({i,j,k,i',j',k'})
\end{equation}
where $\Lambda(i,j,k,i',j',k')$ is given by
\begin{equation}
\Lambda(i,j,k,i',j',k')=\sum_{\boldsymbol{\kappa'}}(-1)^{f(\kappa_A')|i-i'|+f(\kappa_B')|j-j'|+f(\kappa_C')|k-k'|}~p(\boldsymbol{\kappa'}|\pmb{\mathbb{P}},\gamma_p)
\end{equation}
where $f$ is a function such that $f(\kappa_i=-1)=0$ and $f(\kappa_i=1)=1$. We therefore have all the components of $\hat{\rho}_{\mathfrak{E}|\pmb{\mathbb{P}},\gamma_p}$ from which we may numerically find the eigenvalues and compute the first term in the Holevo bound. For the second term in the Holevo bound we need Eve's state conditioned on $\kappa_{A}$. If $\kappa_{A}'=-1$:
\begin{equation}
\hat{\rho}_{\mathfrak{E}|\kappa_{A}'=-1,\pmb{\mathbb{P}}}=\ket{\mathfrak{E}_{\kappa_{A}'=-1}^{\pmb{\mathbb{P}},\gamma_p}}\bra{\mathfrak{E}_{\kappa_{A}'=-1}^{\pmb{\mathbb{P}},\gamma_p}} \otimes\bigg(\sum_{\kappa_B'\kappa_C'} ~ p(\kappa_B',\kappa_C'|\kappa_A'=-1,\pmb{\mathbb{P}},\gamma_p)\ket{\mathfrak{E}_{\kappa_B'\kappa_C'|\kappa_A'=-1}^{\pmb{\mathbb{P}},\gamma_p}}\bra{\mathfrak{E}_{\kappa_B'\kappa_C'|\kappa_A'=-1}^{\pmb{\mathbb{P}},\gamma_p}} \bigg);
\end{equation}
if $\kappa_{A}'=1$:
\begin{equation}
\hat{\rho}_{\mathfrak{E}|\kappa_{A}'=1,\pmb{\mathbb{P}}}=\ket{\mathfrak{E}_{\kappa_{A}'=1}^{\pmb{\mathbb{P}},\gamma_p}}\bra{\mathfrak{E}_{\kappa_{A}'=1}^{\pmb{\mathbb{P}},\gamma_p}} \otimes\bigg(\sum_{\kappa_B'\kappa_C'} ~ p(\kappa_B',\kappa_C'|\kappa_A'=1,\pmb{\mathbb{P}},\gamma_p)\ket{\mathfrak{E}_{\kappa_B'\kappa_C'|\kappa_A'=1}^{\pmb{\mathbb{P}},\gamma_p}}\bra{\mathfrak{E}_{\kappa_B'\kappa_C'|\kappa_A'=1}^{\pmb{\mathbb{P}},\gamma_p}} \bigg).
\end{equation}
The same method explained above may be used to determine components of these density matrices in the $\{\ket{\Phi_{0}^{(B)}},\ket{\Phi_{1}^{(B)}}\}\otimes \{\ket{\Phi_{0}^{(C)}},\ket{\Phi_{1}^{(C)}}\}$ basis. The eigenvalues may then be used to calculate the second term in the Holevo bound.

\subsection{Postselection}
We now demonstrate how the single point quantities may be used to calculate the postselected rate $R_{PS}$. The mutual information $I_{AB}$ may be found by integrating the single point mutual information $\tilde{I}_{AB}$

\begin{equation}
I_{AB}=\int p(\pmb{\mathbb{P}},\gamma_p) \ \tilde{I}_{AB}(\pmb{\mathbb{P}},\gamma_p)  \ d\pmb{\mathbb{P}}~d\gamma_p
\end{equation}
Similarly we do the same for the Holevo information:
\begin{equation}
\chi=\int p(\pmb{\mathbb{P}},\gamma_p) \ \tilde{\chi}(\pmb{\mathbb{P}},\gamma_p)  \ d\pmb{\mathbb{P}}~ d\gamma_p
\end{equation}
By defining the single point rate as $\tilde{R}=\tilde{I}_{AB}-\tilde{\chi}$. Thus the overall rate becomes:
\begin{equation}
R=\int p(\pmb{\mathbb{P}},\gamma_p) \ \tilde{R}(\pmb{\mathbb{P}},\gamma_p)  \ d\pmb{\mathbb{P}}~ d\gamma_p .
\end{equation}
The postselection ensures the parties only use instances of the protocol where the single point rate is positive. Hence the postselected rate $R_{PS}$ becomes:
\begin{align}
R_{PS}=\int p(\pmb{\mathbb{P}},\gamma_p) \ \mathrm{max}\big[\tilde{R}(\pmb{\mathbb{P}},\gamma_p),0\big]  \ d\pmb{\mathbb{P}}~ d\gamma_p \label{eq:integral} \\ 
=\int_{\Gamma} p(\pmb{\mathbb{P}},\gamma_p) \ \tilde{R}(\pmb{\mathbb{P}},\gamma_p)  \ d\pmb{\mathbb{P}}~ d\gamma_p
\end{align}
where $\Gamma$ denotes the region in which the single point rate is positive.
\section{Results}
\label{sec:results}
We now present the numerical results for the post-selected rate of the protocol. By utilising the relation $\tau=10^{-\gamma d}$ and setting $\gamma=0.02/ \mathrm{km}$ (equivalent to $0.2 \mathrm{db / km}$), which  corresponds to state of the art fibre optics, the rate of the protocol is expressed in terms of distances $(d)$ of the parties from the detector. In particular, we consider the symmetric configuration in which each of the parties is located the same distance from the detector. Other asymmetric configurations can be considered within the same framework, by mapping the distance of the user furthest away into the transmissivity of each incoming channel. Thus the results presented here represent the worst case scenario for any other asymmetric configuration of the parties.  

\begin{figure}[ht]
    \centering
    \includegraphics[width=0.5\linewidth]{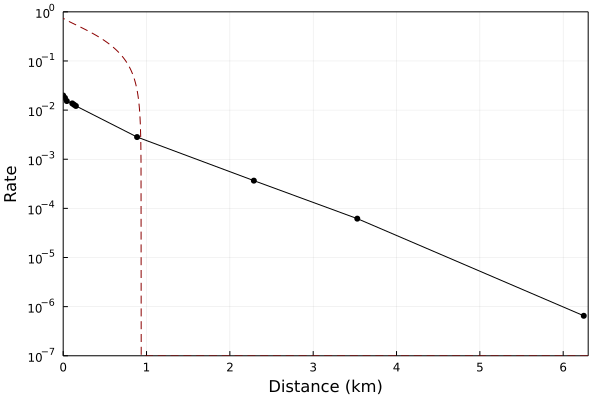}
    \caption{Post-selected rate of the protocol for the symmetric party configuration. Rate plotted with perfect detector efficiency and the variance in all prepared quadratures satisfy $\sigma_A=\sigma_B=\sigma_C=1$. The rate of the equivalent 3-party protocol from \cite{Ottaviani2019} with optimised parameters, under a pure loss attack from is shown for comparison (red dashed line).}
    \label{fig:results}
\end{figure}

Fig.~\ref{fig:results} shows the rate-distance performance of the protocol in the asymptotic limit, assuming that a pure-loss attack is undertaken by Eve. We work with perfect detector efficiency and with the variance of each prepared quadrature $\sigma_{A}=\sigma_{B}=\sigma_{B}=1$ . We note that in general it may be possible to optimise the performance of the protocol over these parameters. Our results demonstrate that a positive rate can be maintained over a greater distance than in the corresponding $3$-party case (shown for comparison in Fig \ref{fig:results}, albeit at the cost of lower rates at short distances). In particular the new protocol outperforms the equivalent protocol without postselection for distances greater than $\sim 1\mathrm{km}$.

\section{Conclusion}
We have demonstrated a 3-party CV-MDI-QKD protocol that combines a generalised Bell detection with a postselection regime based on performing reconciliation on the signs of prepared quadratures of coherent states. We show that improved rate-distance performance is possible compared to the equivalent $3$-party protocol without postselection, allowing a rate in excess of $10^{-4}$ bits per use at greater than $3\mathrm{km}$  and a positive rate for distances of up to $\sim 6\mathrm{km}$. Our protocol also outperforms the equivalent protocol without postselection for distances greater than $\sim 1\mathrm{km}$. Moreover since these protocols have exactly the same structure in terms of state preparation and the detector relay, it is possible to use one such relay to perform either protocol, choosing whichever will give the higher rate. That is, if the users are able to establish their distances from the detector, they choose whether or not to announce the absolute values of their quadratures and undertake postselection depending on whether or not this will produce a better rate. Whilst $\sigma_A,\sigma_B,\sigma_C$ are preset so any optimisation over these parameters must consider both protocols simultaneously it is still possible to retain the advantages of higher rate at shorter distances from the non-postselected protocol in addition to the improved long distance performance from our protocol. 

The need to undertake a high-dimensional numerical integral, given in Eq.~(\ref{eq:integral}) to compute the post-selected key rate, limits our analysis to the $3$-party case and pure-loss attacks. Nonetheless it may be possible to extend the study to the general $N$ party case, maintaining the same structure of detector as in \cite{Ottaviani2019} and considering entangling cloner attacks. Thus, our new protocol demonstrates that secure, multi-party conferencing can be achieved over improved distances, while retaining the security advantages of an MDI QKD protocol.  

\label{sec:conclusion}

\acknowledgements
A.I.F. acknowledges funding from the EPSRC via a Doctoral Training Partnership (EP/R513386/1). S. P. acknowledges funding from 
the European Union via the flagship project \textquotedblleft Continuous Variable Quantum Communications\textquotedblright\ (CiViQ, Grant agreement No. 820466) and  
the EPSRC via the UK Quantum Communications Hub (Grant No. EP/T001011/1). The authors would like to thank Kieran Wilkinson for helpful discussions. 
\newpage
\bibliography{MDI}

\end{document}